# Superconductivity in higher titanium oxides

K. Yoshimatsu[1,*], O. Sakata[2,3], and A. Ohtomo[1,3,*]

[1]*Department of Chemical Science and Engineering, Tokyo Institute of Technology, 2-12-1 Ookayama, Meguro-ku, Tokyo 152-8552, Japan*
[2]*Synchrotron X-ray Station at SPring-8, National Institute for Materials Science (NIMS), Sayo, Hyogo 679-5148, Japan*
[3]*Materials Research Centre for Element Strategy (MCES), Tokyo Institute of Technology, Yokohama 226-8503, Japan*

*Author to whom correspondence should be addressed; Electronic mail: k-yoshi@apc.titech.ac.jp & aohtomo@apc.titech.ac.jp

**Abstract**

After the discovery of zero resistivity in mercury at the temperature of liquid helium, several conventional superconductors have been identified from single elements and alloys over the course of decades[1-3]. The Bardeen–Cooper–Schrieffer (BCS) theory explains the unique properties of superconductors[4] and guides materials design for achieving a higher superconducting transition temperature ($T_c$) under the criterion of the large electron–phonon coupling constant ($\lambda_{ep}$). Magnesium diboride with $T_c$ = 39 K has been discovered using this strategy[5]. A further increase in $\lambda_{ep}$ is ineffective in enhancing $T_c$ because the localization of electron pairs in real space results in the formation of a charge density wave, and the highest $T_c$ considered within the BCS theory is approximately 40 K. This BCS wall has been easily overcome with the demonstration of perovskite-type copper oxides[6]. Unconventional materials, including fullerenes[7] and iron pnictides[8], have reigned over the class of high–$T_c$ superconductors. However, a conventional superconductor (sulphur hydride) recently created a new record of $T_c$ = 203 K under an ultra–high pressure[9]. Such old materials but new superconductors are again attracting significant attention[10]. Here, we report new superconductors produced from higher titanium oxides, which are $Ti_4O_7$ and $\gamma$-$Ti_3O_5$ films synthesized using advanced epitaxial growth on specific oxide substrates. Their $T_c$ are 3.0 and 7.1 K, respectively. The latter is the highest known value among simple oxides. The temperature dependence of resistivity strongly depended on the growth atmosphere. Higher titanium-oxide films grown under a more oxidative condition exhibited metal–insulator transition accompanied by clear hysteresis at ~150 K. The insulating phase was suppressed in the films grown under a less oxidative condition, and the superconducting phase appeared at low temperatures. These results and the previous theoretical prediction suggest that oxygen non-stoichiometry and epitaxial stabilization play key roles in the realization of bipolaronic superconductivity in these titanium oxides.

**Introduction**

Titanium oxides are one of the most popular compounds among simple oxides, *e.g.* $TiO_2$, which acts as a photocatalyst for water-oxidation reactions under the illumination of ultraviolet light[11]. Their abundance is also useful in applications of solar cells combined with visible-light-absorbing molecular dyes[12]. Except for the fully oxidized compound, titanium oxides have partially filled $d$ states and their exotic properties have captured attention: superconductivity exhibited by the monoxide (TiO)[13], Peierls transition exhibited by the sesquioxide ($Ti_2O_3$)[14], and metal–insulator transition (MIT) exhibited by the Magnéli phase ($Ti_nO_{2n-1}$, $n \geq 4$)[15-19]. In the periodic table, titanium oxides are the first group of simple oxides indicating metallicity, and all the simple oxides of scandium or much lighter elements are insulating. Therefore, the choice of titanium oxides is favourable for large electron–phonon coupling.

The Magnéli phase exhibits unique low-dimensional structures characterized by shear planes. These shear planes correspond to the rutile $TiO_2$ (121) planes and amputate the edge-shared infinite $TiO_6$ chains at every $n$ $TiO_6$ blocks with shifting by a half of the unit cell. Figure 1a shows a schematic of the crystal structure for the first member of Magnéli-phase $Ti_4O_7$.

Trititanium pentoxide ($Ti_3O_5$) with polymorphisms ($\alpha$-, $\beta$-, $\gamma$-, $\delta$-, and $\lambda$-phases) is a neighbour of the Magnéli phase[20-24]. In contrast to the Magnéli phase, there are no shear planes, as illustrated in Fig. 1b. Because of difficulty in the growth of a single crystal due to polymorphism, their physical properties are still under debate. Several studies have dealt with the structural phase transitions accompanying MIT, which are induced under the specific conditions ($\alpha \leftrightarrow \beta$ at 450 K[20], $\delta \leftrightarrow \gamma$ at 240 K[21,24], and $\beta \leftrightarrow \lambda$ by irradiation using visible-light pulses[23]).

This paper reports the discovery of superconductivity for $Ti_4O_7$ and $\gamma$-$Ti_3O_5$ in a thin film form. The higher titanium oxides join in a class of simple-oxide superconductors, and $\gamma$-$Ti_3O_5$ now holds the highest $T_c$ among them. The possible mechanism behind the superconductivity is discussed on the basis of electrical measurements and theoretical predictions. We conclude that bipolaronic superconductivity is realized for both of the higher titanium oxides with the assistance of oxygen non-stoichiometry and epitaxial stabilization.

**Results**

The formation of higher titanium-oxide films was verified using x-ray diffraction (XRD). The out-of-plane XRD patterns showed intense reflections from the $Ti_4O_7$ films grown on $(LaAlO_3)_{0.3}$–$(SrAl_{0.5}Ta_{0.5}O_3)_{0.7}$ (LSAT) (100) substrates and the $\gamma$-$Ti_3O_5$ film grown on $\alpha$-$Al_2O_3$ (0001) substrates (Figs. 2a and 2b, respectively). These substrates are insulating, non-magnetic, and exhibit high reduction resistance, providing advantages in the growth and search of a superconducting sample. However, because titanium oxides form various polymorphisms with different ratios of oxygen to titanium, their crystal structures must be carefully controlled and distinguished. Pulsed-laser deposition (PLD) with a $TiO_x$ target ($x \sim 1.5$) was conducted at a high temperature to tune the degree of subtle oxidation by feeding a small amount of oxygen gas or even flowing inert Ar gas. Then, we used the tilt angle $\chi$-dependence of $2\theta$-$\theta$ XRD profiles to survey the asymmetric film reflections (see Figs. S1 and S5 in Supplemental Materials). Reflections coming from the substrate and film were found at characteristic $\chi$ angles. Since the intensities of the film reflections were too weak to determine the $d$ values of interplanar spacing distances precisely, synchrotron radiation XRD measurements were also performed (see Figs. S2, S3, and S6). From the $d$ values and $\chi$ angles, we identified the Miller indices as those listed in Tables S1 and S2. In comparison to the previous structural analyses of higher titanium oxides[15,21-24], we concluded that the films grown on LSAT (100) and $\alpha$-$Al_2O_3$ (0001) substrates were $Ti_4O_7$ and

$\gamma$-Ti$_3$O$_5$, respectively.

The in-plane epitaxial relationship between the substrates and films were also investigated. The Ti$_4$O$_7$ 134 reflection appeared every 90° in the XRD azimuth $\phi$-scan around the film normal, indicating four-fold rotational domains (Fig. S4). The in-plane (out-of-plane) epitaxial relationships between the film and substrate were Ti$_4$O$_7$ [010] // LSAT [010], [001] and Ti$_4$O$_7$ [0-10] // LSAT [010], [001] (Ti$_4$O$_7$ [101] // LSAT [100]). The $\gamma$-Ti$_3$O$_5$ 143 reflection appeared every 60° in the XRD azimuth $\phi$-scan around the film normal (Fig. S7). The in-plane (out-of-plane) epitaxial relationships were $\gamma$-Ti$_3$O$_5$ [100] // $\alpha$-Al$_2$O$_3$ [10–10], [01–10], [–1100] and $\gamma$-Ti$_3$O$_5$ [-100] // $\alpha$-Al$_2$O$_3$ [10–10], [01–10], [–1100] ($\gamma$-Ti$_3$O$_5$ [011] // $\alpha$-Al$_2$O$_3$ [0001]). Six-fold rotational domains of the orthorhombic films were found on the trigonal substrates.

The electrical properties of the films were investigated using the temperature dependence of resistivity (Fig. 3). The resistivity curves strongly depended on the growth atmosphere for Ti$_4$O$_7$ films (Fig. 3a). For the film grown under $P_{O2}$ = 1 × 10$^{-7}$ Torr, MIT accompanied by clear hysteresis was found at around 150 K, which is in agreement with the behaviour of a bipolaron insulator of bulk Ti$_4$O$_7$[16–18]. In contrast, the insulating behaviours were strongly suppressed for the film grown under $P_{Ar}$ = 1 × 10$^{-3}$ Torr; the upturn in resistivity was weak and no hysteresis was found over the temperature range. If we account for the lower degree of oxidation at $P_{Ar}$ = 1 × 10$^{-3}$ Torr, oxygen deficiency may be responsible for the suppression of the insulating states. Furthermore, superconductivity was observed at low temperatures. The Ti$_4$O$_7$ film grown under an intermediate condition ($P_{Ar}$ = 1 × 10$^{-6}$ Torr) exhibited both hysteresis and superconducting characteristics in the resistivity curve. We will refer to the Ti$_4$O$_7$ films grown under $P_{O2}$ = 1 × 10$^{-7}$ Torr ($P_{Ar}$ = 1 × 10$^{-3}$ Torr) as insulating (superconducting) ones in the following discussion.

The variation in the Hall coefficient ($R_H$) during warming exhibited a tendency similar to that of resistivity. For the insulating Ti$_4$O$_7$ film, the temperature dependence of the inverse $R_H$ (inset of Fig. 3a) suddenly decreased at around 150 K, suggesting that the MIT was induced by the depletion of hole carriers. The inverse $R_H$ at 10 K was four orders of magnitude smaller than that at 300 K. The MIT in the bulk is associated with the formation of bipolarons[16–18], which remains robust in the insulating Ti$_4$O$_7$ film at low temperatures. In contrast, the inverse $R_H$ for the superconducting Ti$_4$O$_7$ film was independent of temperatures, and even the value at 10 K was comparable to that at 300 K, suggesting the suppression of a bipolaronic insulating state.

The temperature dependence of the resistivity for the $\gamma$-Ti$_3$O$_5$ film exhibited a complex curve along three electronic phase transitions: MIT around 350 K, insulator–insulator transition around 100 K, and superconducting transition (Fig. 3b). The observed transitions have not been reported for bulk $\gamma$-Ti$_3$O$_5$[20–24], nor are they reminiscent of a structural phase transition to $\delta$-Ti$_3$O$_5$ at ~240 K[23] accompanying MIT[24]. The intermediate transition may be related to the MIT of Ti$_4$O$_7$ due to their similar transition temperatures. Nevertheless, the resistivity upturn was much weaker, suggesting the suppression of the insulating states, as with the case of the superconducting $Ti_4O_7$ film. The sign and magnitude of the $R_H$ also reflected this correspondence (inset of Fig. 3(b)). From these viewpoints, the superconductivities of the two different higher titanium oxides likely have the same origin.

The temperature dependence of resistivity around the temperature of liquid helium indicates further similarity between the superconducting $Ti_4O_7$ and $\gamma$-$Ti_3O_5$ films (Figs. 4a and 4b, respectively). The $T_{c,\,onset}$ of $Ti_4O_7$ and $\gamma$-$Ti_3O_5$ were 3.0 K and 7.1 K, respectively. Note that the $T_c$ of both films exceeded that of other simple-oxide superconductors (TiO ($T_c$ = 2.3 K), NbO ($T_c$ ~1.4 K), and SnO ($T_c$ = 1.4 K under 9.3 GPa))[13,25,26]. In contrast, the magnetic field dependence of the superconducting transitions exhibited a difference. The superconductivity in the $Ti_4O_7$ film disappeared above 2 K under a magnetic field of 7 T. As for the $\gamma$-$Ti_3O_5$ films, superconductivity remained robust even under 9 T. In addition, a clear Meissner effect was observed for the $\gamma$-$Ti_3O_5$ film (inset of Fig. 4b). The diamagnetic signals were analysed with an equation applicable to a uniformly magnetized ellipsoid[27]. The superconducting volume fraction was estimated to be larger than >90% at 3.3 K. These results satisfactorily verify the superconducting phase residing uniformly in the higher titanium-oxide films.

**Discussion**

Finally, we will discuss the possible mechanism behind the superconductivity of the higher titanium oxides. The MIT of the stoichiometric $Ti_4O_7$ bulk is based on the premise of the bipolaronic interaction[16–19]. Sharp increase in resistivity and hysteresis at the MIT are strong evidence for the bipolaron formation[16–18]. The insulating $Ti_4O_7$ film exhibiting such characteristics can be regarded as a bipolaronic insulator at low temperatures. Chakraverty *et al.* predicted that a bipolaronic insulator (or charge density wave (CDW) insulator) undergoes a transition to a superconductor when $\lambda_{ep}$ is large[28–30]. They also suggested $Ti_4O_7$ with the largest value of $\lambda_{ep} = 3$ as a candidate material[30]. Therefore, experimental verifications for superconductivity in $Ti_4O_7$ were attempted by applying high pressures. However, no superconducting transition was observed under a hydrostatic pressure of up to 5.0 GPa, although the high-temperature metallic phase was extended down to 3 K[17,18].

For a bipolaronic system, the bipolaron density is a key parameter in the electronic phase diagram[29]. De Jongh described in Ref. 29 that 'the concentration of bipolaron is nearly always too high, a metal–insulator transition occurs at low temperatures (as in $Ti_4O_7$) accompanied by a CDW and a structural change, instead of the much-sought-for transition to superconductivity'. Our growth of $Ti_4O_7$ films under Ar gas atmosphere aims at inducing extra Ti 3*d* electrons by oxygen vacancies which dilute the bipolaron density. In fact, the inverse $R_H$ of the superconducting film suggests the suppression of the bipolaron formation (Fig. 3(a)). In the bipolaronic phase diagram shown in Fig. 5 of Ref. 29, the bipolaronic insulator and superconductivity co-exist at a specific bipolaron density. Our $Ti_4O_7$ film grown under a moderate growth atmosphere ($P_{Ar} = 1 \times 10^{-6}$ Torr) exhibited hysteresis at ~150 K and superconductivity at 2.9 K (see Fig. 3a and Fig. S9). This result provides further evidence for the bipolaronic superconductivity.

$Ti_4O_7$ films grown on $MgAl_2O_4$ (100) substrates also exhibited superconductivity (Fig. S10). Thus, the observed superconductivity is intrinsic to the $Ti_4O_7$ phase. Furthermore, superconductors composed of Mg, Al, Ti, and O with $T_c$ of more than 3 K are not yet known, indicating that any elements from the substrates cannot induce the superconductivity in our samples.

This first observation of superconductivity in a $Ti_4O_7$ film demonstrates the importance of the epitaxial thin film. Titanium-based simple oxides with various chemical formulae[30] and polymorphisms easily transform from one to another, and subtle tuning of oxygen stoichiometry causes modulation of carrier density. Epitaxial growth on LSAT substrates enables us to stabilize the Magnéli phase. In fact, the $\gamma$-$Ti_3O_5$ and $Ti_4O_7$ films can also be grown on different substrates under the same growth atmosphere ($P_{o2} = 1 \times 10^{-7}$ Torr). The lack of these advantages may be inevitable for hidden superconducting phases in bulk specimens. We note that recent reports on the enhancement of $T_c$ in FeSe are also realized in a form of thin film[31].

For bulk $\gamma$-$Ti_3O_5$, the formation of bipolarons has not been reported because the MIT of other Magnéli-phase titanium oxides occurs at temperatures lower than the structural phase transition at ~ 240 K[18]. There was no sign of such a structural phase transition in the resistivity curve of the $\gamma$-Ti$_3$O$_5$ film (Fig. 3b), suggesting that the metallic $\gamma$-phase was stabilized in the form of an epitaxial thin film. The first-principle calculations revealed a one-dimensional conducting pathway along the *c*-axis arising from the density of states at the Fermi level[24]. The low-dimensional electronic structure may lead to the pairing of electrons (bipolaron) in $\gamma$-Ti$_3$O$_5$ at a low temperature. On the other hand, the small number of studies on $\gamma$-Ti$_3$O$_5$ makes it difficult to discuss the strength of the electron–phonon interaction, the formation of bipolarons, and the density of states at the Fermi level. Nevertheless, given the close relationship between Ti$_4$O$_7$ and $\gamma$-Ti$_3$O$_5$, it is natural for us to presume a similar origin of (bi)polaronic superconductivity.

**Experiments**

A TiO$_x$ ceramic tablet with the majority phase of Ti$_2$O$_3$ was prepared using a conventional solid-state reaction method. Ti (3N) and TiO$_2$ (4N) powders with a molar ratio of 1:3 were mixed and pressed into a pellet. This was sintered at 1000°C for 12 $h$ in vacuum. Prior to the film growth, LSAT and α-Al$_2$O$_3$ substrates were annealed in air to obtain a step-and-terrace surface. The annealing conditions were 1200°C for 3 $h$ for LSAT, and 1100°C for 3 $h$ for α-Al$_2$O$_3$. The higher titanium-oxide films were grown using PLD in an ultra-high-vacuum chamber. KrF excimer laser pulses (5 Hz, 2.0 J/cm$^2$) were focused on the TiO$_x$ ceramics tablets. The growth temperature was set at 900°C. The chamber pressure was controlled with the continuous flow of oxygen or Ar gas (6N purity for both). Introduction of oxygen (Ar) gas during the growth corresponds to oxidation (reduction) of the films. After the growth, the gas flow was stopped immediately, and the samples were quenched to room temperature.

The crystal structures of the films were characterized using XRD with Cu K$\alpha_1$ radiation (Rigaku, SmartLab) and synchrotron radiation at BL15XU in SPring-8. The photon energy of the synchrotron radiation was set at 15 keV ($\lambda$ = 0.826 Å). The temperature dependence of resistivity was measured using a standard four-probe method with a physical properties measurement system (Quantum Design, PPMS). The temperature dependence of the Hall measurements was also measured using PPMS in a standard six-terminal geometry. The temperature dependence of magnetization was measured using magnetic properties measurement system (Quantum Design, MPMS).

**Acknowledgement**

The authors greatly thank Prof. S. Saito, Dr. S. Okamoto, Dr. T. Koretsune, and Dr. R. Arita for their useful discussion. The authors also thank M. Azuma and H. Hojo for assistance of magnetization measurements. The Synchrotron XRD measurements were performed under the approval of the NIMS Synchrotron x-ray station at SPring-8 (Proposal No. 2015B4700). The authors are also grateful to K. Yokoyama, S. Takeda, K. Kagoshima, and J. Matsui, University of Hyogo, for their technical contribution and to Y. Shimada and Y. Katsuya for their technical support. This work was partly supported by MEXT Elements Strategy Initiative to Form Core Research Centre and a Grant-in-Aid for Scientific Research (Nos. 15H03881 and 16H05983) from the Japan Society for the Promotion of Science Foundation.

**Author contributions**

K. Y. performed all of the experiments, interpreted the data, and wrote the manuscript. O. S. managed the synchrotron radiation XRD measurements. A. O. supervised the whole research.

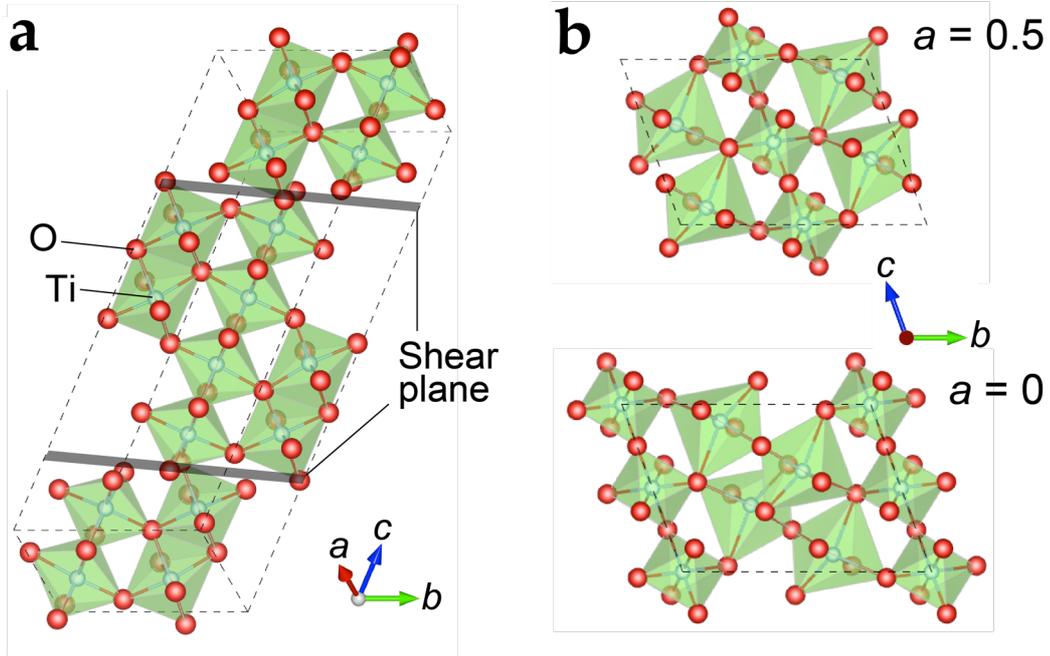

**Figure 1 Crystal structures of Ti$_4$O$_7$ and γ-Ti$_3$O$_5$. a**, The first member of Magnéli-phase Ti$_n$O$_{2n-1}$ ($n$ = 4)[15–19] with a triclinic cell ($a$ = 5.597 Å, $b$ = 7.125 Å, $c$ = 20.429 Å, $\alpha$ = 67.7°, $\beta$ = 57.16°, $\gamma$ = 108.76°)[15,16,19]. The shear planes, corresponding to the (121) planes of the rutile-type TiO$_2$, amputate the edge-shared TiO$_6$ chains at every four TiO$_6$ blocks. A group of TiO$_6$ blocks shifts along the shear plane by half of the spacing between each chain. In the nominal composition, a TiO$_6$ tetramer has two electrons occupying the Ti 3$d$ states. **b**, γ-Ti$_3$O$_5$ is one of the five polymorphs, α-, β-, γ-, δ-, and λ-phases[20–24] with a monoclinic cell ($a$ = 5.0747 Å, $b$ = 9.9701 Å, $c$ = 7.1810 Å, $\alpha$ = 109.865°)[21]. In contrast to Ti$_4$O$_7$, there is no shear plane, and so Ti$_3$O$_5$ is out of the Magnéli phase. However, since the chemical formula is consistent with that of the Magnéli phase (Ti$_n$O$_{2n-1}$ at $n$ = 3), it is sometimes designated as the first member of the Magnéli phase.

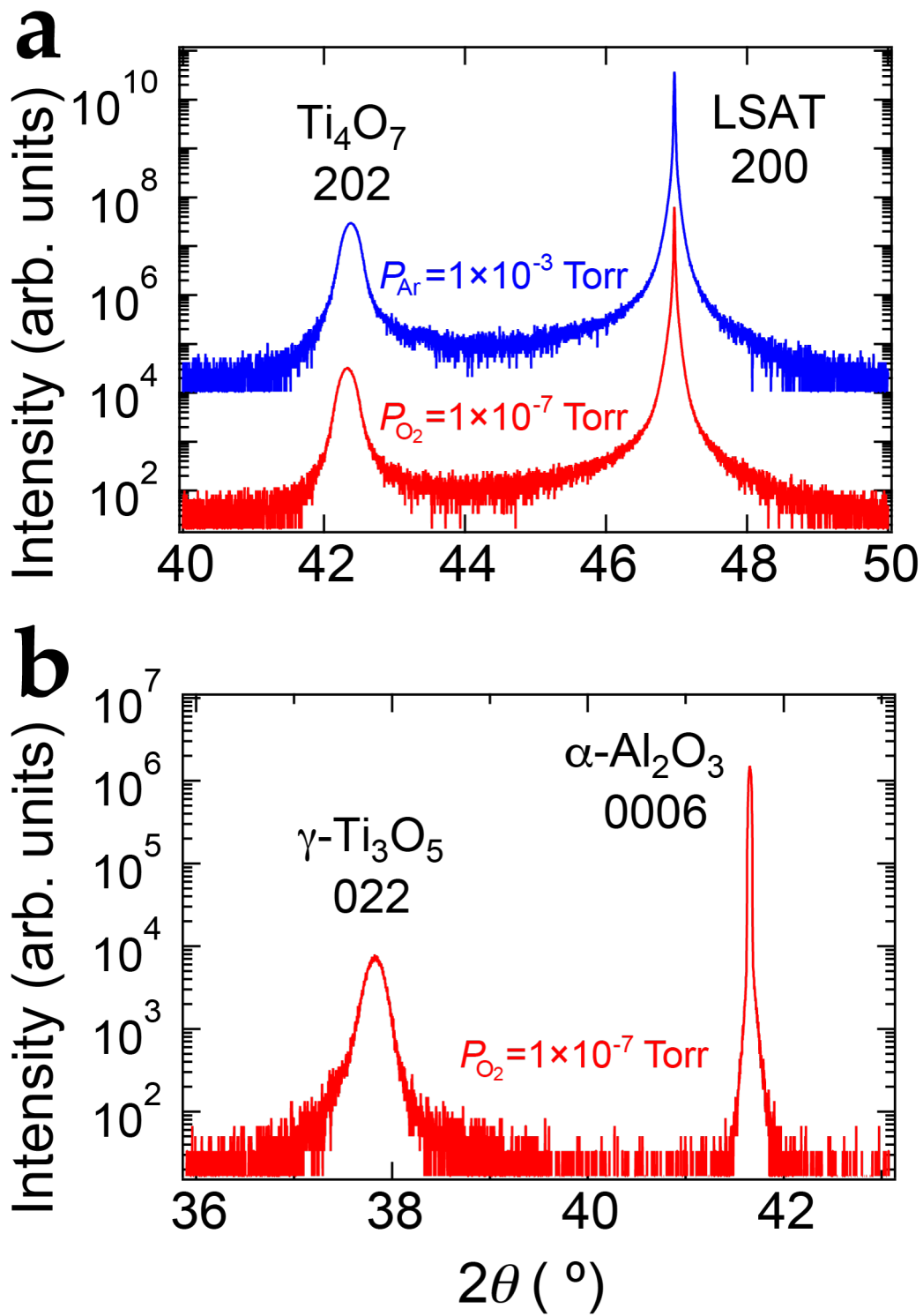

**Figure 2 XRD patterns of $Ti_4O_7$ and $\gamma$-$Ti_3O_5$ films. a,** The out-of-plane XRD patterns for $Ti_4O_7$ films grown on LSAT (100) substrates in Ar gas at $1 \times 10^{-3}$ Torr

(top) and in oxygen gas at 1 × 10$^{-7}$ Torr (bottom). Irrespective of the growth condition, Ti$_4$O$_7$ 202 reflection was detected at 2$\theta$ = 42.38°, corresponding to $d_{202}$ = 2.13 Å. No other film reflections except for the 404 reflection at 2$\theta$ = 92.60° was detected in wide-range XRD patterns. Synchrotron radiation XRD measurements revealed slight difference in $d$ values between these films (Table S1). **b**, The out-of-plane XRD pattern for the $\gamma$-Ti$_3$O$_5$ film grown on $\alpha$-Al$_2$O$_3$ (0001) substrates in oxygen gas at 1 × 10$^{-7}$ Torr. The $\gamma$-Ti$_3$O$_5$ 022 reflection was detected at 2$\theta$ = 37,83°, corresponding to $d_{022}$ = 2.38 Å. The out-of-plane single orientation was verified using wide-range XRD patterns.

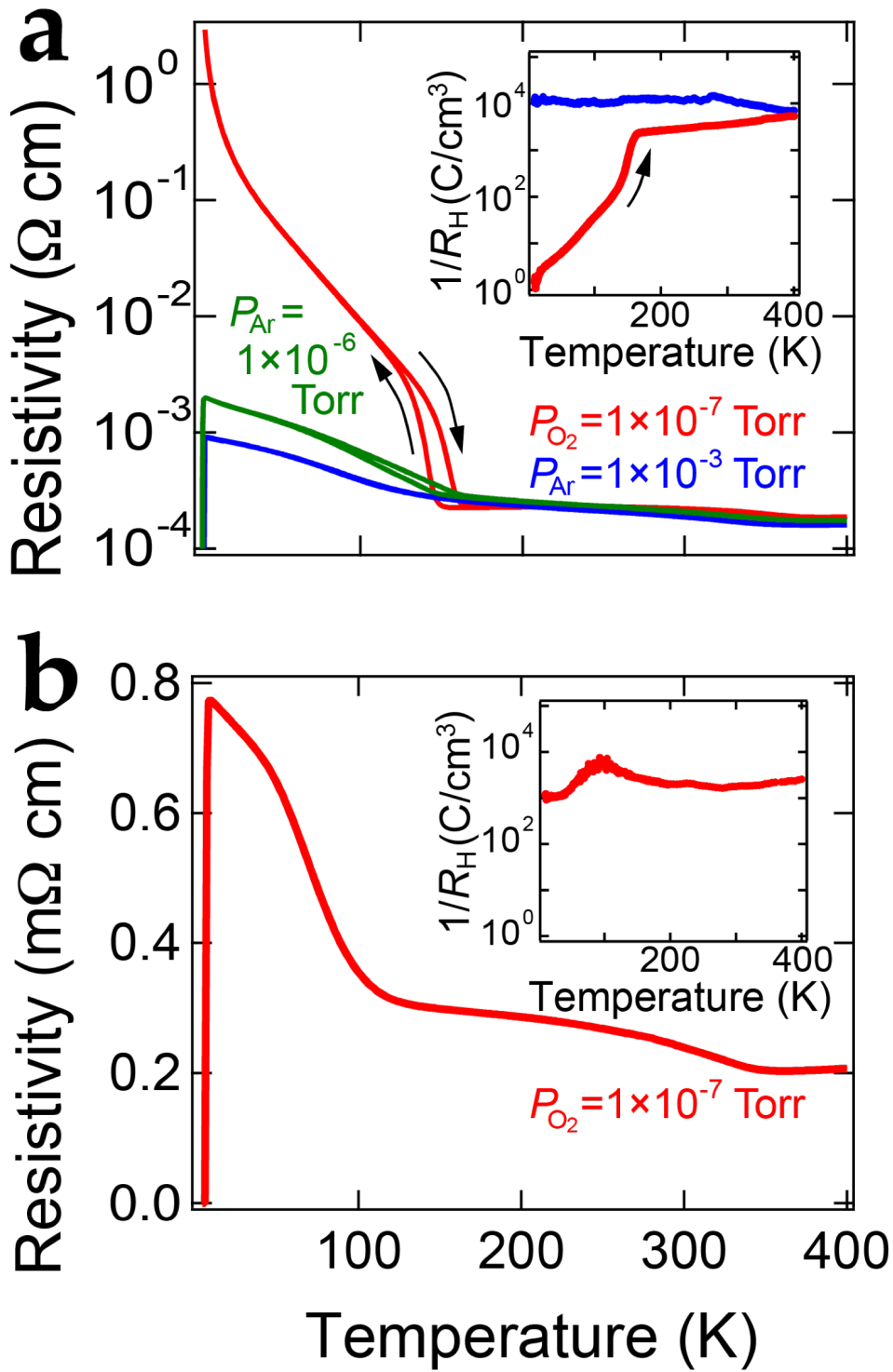

**Figure 3 Temperature dependence of resistivity for Ti$_4$O$_7$ and γ-Ti$_3$O$_5$ films. a,** Temperature dependence of resistivity for Ti$_4$O$_7$ films grown under three

different conditions. The resistivity curves strongly depended on the growth conditions. The inset shows the temperature dependence of the Hall measurement. At 300 K (10 K), the inverse $R_H$ was 3.6 × 10$^3$ (1.5) and 1.2 × 10$^4$ (1.2 × 10$^4$) C/cm$^3$ for the films grown under $P_{O2}$ = 1 × 10$^{-7}$ Torr and $P_{Ar}$ = 1 × 10$^{-3}$ Torr, respectively. The sudden decrease in the inverse $R_H$ only occurred at around 150 K for the former. **b,** Temperature dependence of resistivity for the $\gamma$-Ti$_3$O$_5$ film. The inset shows the temperature dependence of the Hall measurement. The inverse $R_H$ almost remained the same (~10$^3$ cm$^3$/C) over the entire temperature range.

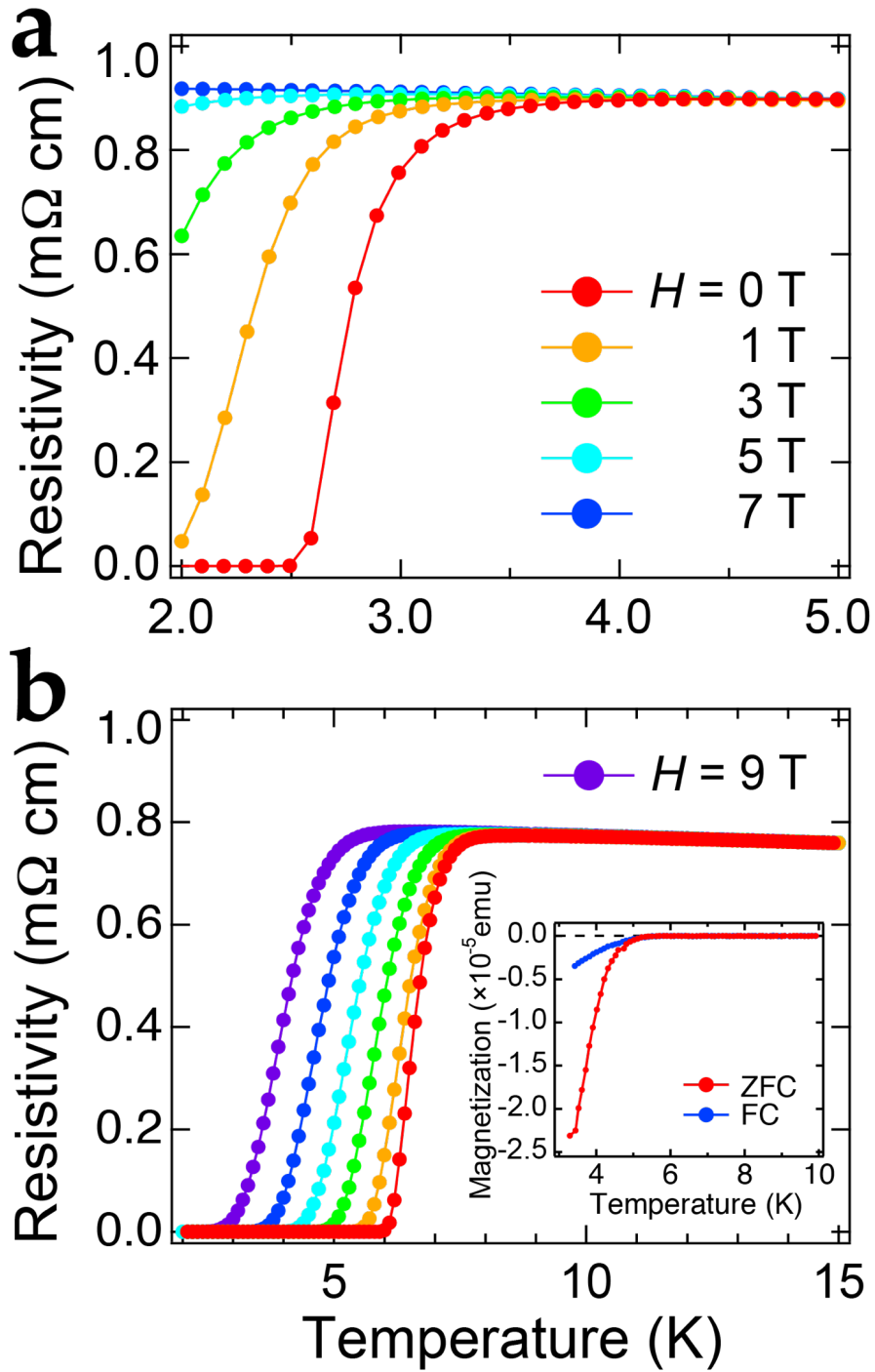

**Figure 4 Superconductivity of Ti$_4$O$_7$ and $\gamma$-Ti$_3$O$_5$ films. a,** Temperature dependence of resistivity of the Ti$_4$O$_7$ film grown under $P_{Ar} = 1 \times 10^{-3}$ Torr at low temperatures. The $T_c$ of the film was estimated to be 3.0 K ($T_{C, onset}$), 2.7 K ($T_{C, mid}$),

and 2.5 K ($T_{C, zero}$), with no applied magnetic fields. The superconducting states were gradually degraded under applied magnetic fields. $T_c$ shifted toward a lower temperature under a higher magnetic field, and the superconducting phase finally disappeared for the Ti$_4$O$_7$ film at above 2 K. **b**, Temperature dependence of resistivity of the γ-Ti$_3$O$_5$ film at low temperatures. The $T_c$ of the film was estimated to be 7.1 K ($T_{C, onset}$), 6.6 K ($T_{C, mid}$), and 5.8 K ($T_{C, zero}$), with no applied magnetic fields. The inset shows temperature dependence of magnetization for the γ-Ti$_3$O$_5$ film at low temperatures. FC and ZFC denote field-cooling and zero-field cooling curves, respectively. Clear diamagnetic signals were observed, indicating the Meissner effect in the film.

Supplemental Materials

# Superconductivity in higher titanium oxides

K. Yoshimatsu[1,*], O. Sakata[2,3], and A. Ohtomo[1,3,*]

[1]*Department of Chemical Science and Engineering, Tokyo Institute of Technology, 2-12-1 Ookayama, Meguro-ku, Tokyo 152-8552, Japan*

[2]*Synchrotron X-ray Station at SPring-8, National Institute for Materials Science (NIMS), Sayo, Hyogo 679-5148, Japan*

[3]*Materials Research Centre for Element Strategy (MCES), Tokyo Institute of Technology, Yokohama 226-8503, Japan*

*Author to whom correspondence should be addressed; Electronic mail: k-yoshi@apc.titech.ac.jp & aohtomo@apc.titech.ac.jp

**X-ray diffraction (XRD) measurements for the $Ti_4O_7$ films**

Figure S1 shows contour map of film- and substrate-reflections intensity, plotted against scattering angle $2\theta$ and tilt angle $\chi$, for the $Ti_4O_7$ films grown under $P_{Ar} = 1 \times 10^{-3}$ Torr (superconducting $Ti_4O_7$). The peak locations corresponding to the film and substrate reflections were verified. Based on this survey, we have performed synchrotron radiation XRD at BL15XU in SPring-8 for both the insulating and superconducting $Ti_4O_7$ films. The measured reflection profiles were shown in Figs. S2 (a–h) (insulating $Ti_4O_7$ film) and Figs. S3 (a–h) (superconducting $Ti_4O_7$ film). Signal to noise ratio was significantly improved using high-flux synchrotron radiation. From the $d$ values of interplanar spacing distances and $\chi$ angles, the Miller indices were assigned as listed in Tables S1. For the $Ti_4O_7$ 134 reflection, the XRD azimuth $\phi$-scans around the film normal were also performed to reveal the rotational domains of the films (Figs. S4). The peaks appeared every 90°, indicating four-fold rotational domains in the films. From these XRD analyses, the in-plane (out-of-plane) epitaxial relationships between the films and substrate were determined to be $Ti_4O_7$ [010] // $(LaAlO_3)_{0.3}$–$(SrAl_{0.5}Ta_{0.5}O_3)_{0.7}$ (LSAT) [010], [001] and $Ti_4O_7$ [0–10] // LSAT [010], [001] ($Ti_4O_7$ [101] // LSAT [100]).

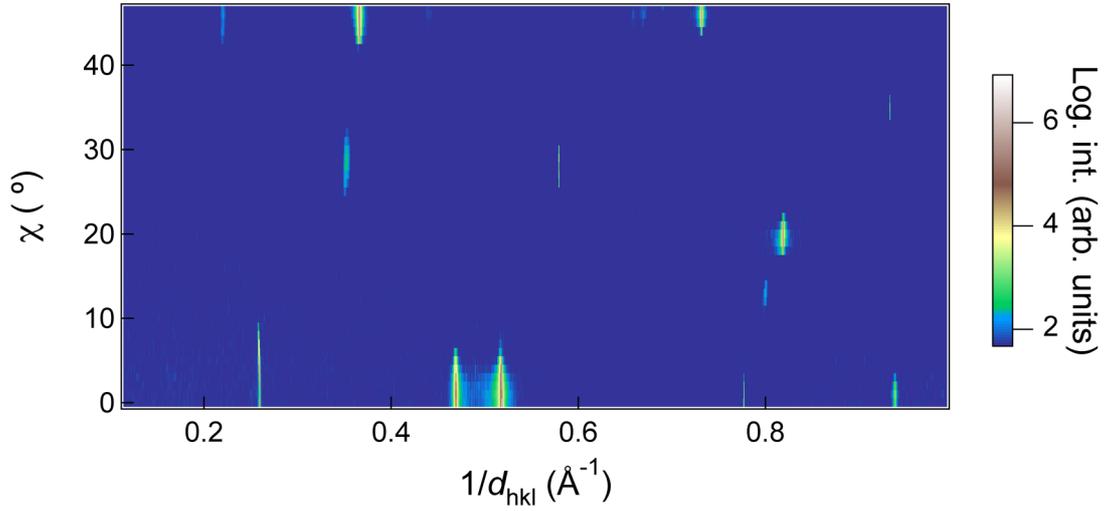

**Figure S1.** Contour map of film- and substrate-reflections intensity, constructed from 2θ-θ profiles measured by stepwisely varying tilt angle χ, for the superconducting $Ti_4O_7$ film grown on the LSAT (100) substrate.

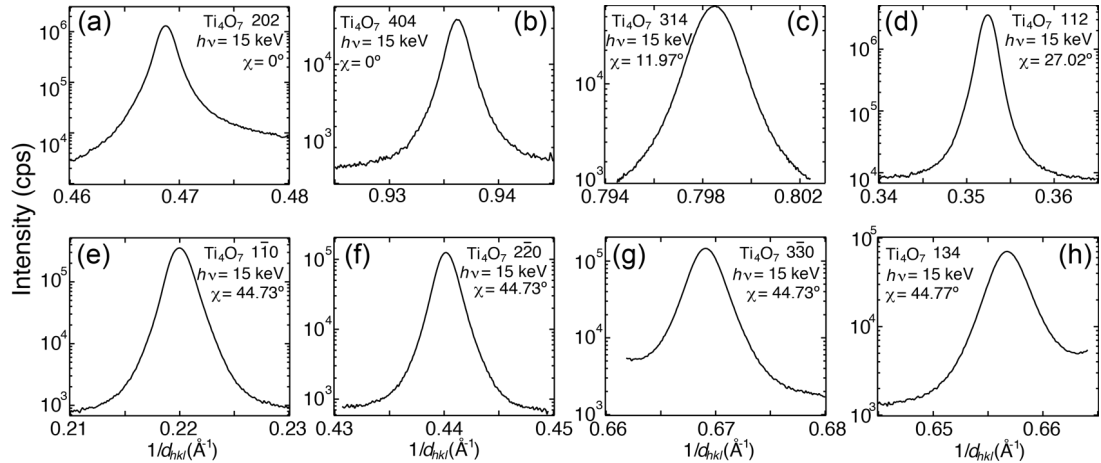

**Figure S2.** Some of the film-reflection profiles measured for the insulating $Ti_4O_7$ film. (a) 202, (b) 404, (c) 314, (d) 112, (e) 1–10, (f) 2–20, (g) 3–30, and (h) 134 reflections.

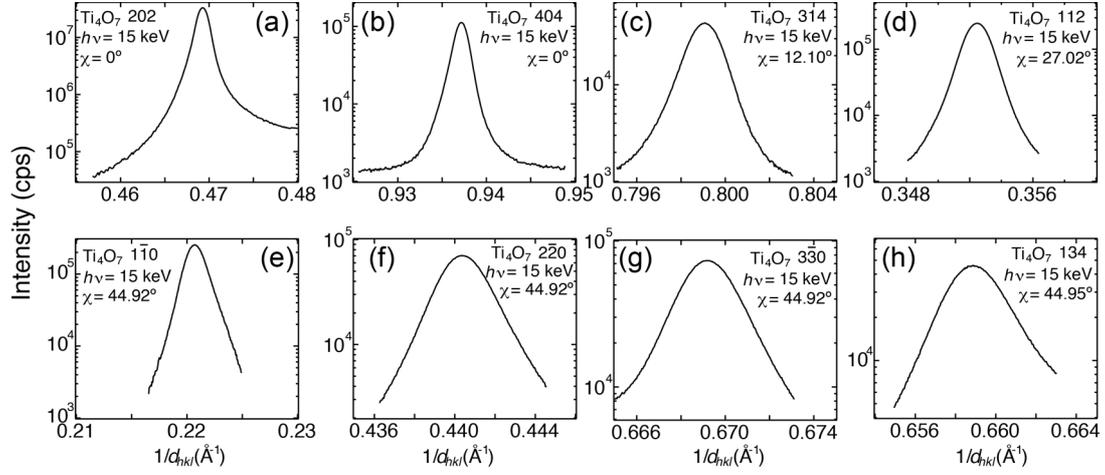

**Figure S3.** Some of the film-reflection profiles measured for the superconducting Ti₄O₇ film. (a) 202, (b) 404, (c) 314, (d) 112, (e) 1–10, (f) 2–20, (g) 3–30, and (h) 134 reflections.

**Table S1.** List of Miller indices, $d$ values of interplanar spacing distances, and tilt angle $\chi$ for the insulating (left) and superconducting (light) Ti₄O₇ films.

| No | hkl | $d_{hkl}$ (Å) | $\chi$ (°) | No | hkl | $d_{hkl}$ (Å) | $\chi$ (°) |
|---|---|---|---|---|---|---|---|
| 1 | 202 | 2.133 | 0 | 1 | 202 | 2.131 | 0 |
| 2 | 404 | 1.068 | 0 | 2 | 404 | 1.067 | 0 |
| 3 | 314 | 1.252 | 11.97 | 3 | 314 | 1.251 | 12.10 |
| 4 | 112 | 2.837 | 27.02 | 4 | 112 | 2.837 | 27.02 |
| 5 | 1–10 | 4.544 | 44.73 | 5 | 1–10 | 4.531 | 44.92 |
| 6 | 2–20 | 2.272 | 44.73 | 6 | 2–20 | 2.274 | 44.92 |
| 7 | 3–30 | 1.494 | 44.73 | 7 | 3–30 | 1.457 | 44.92 |
| 8 | 134 | 1.522 | 44.77 | 8 | 134 | 1.518 | 44.95 |

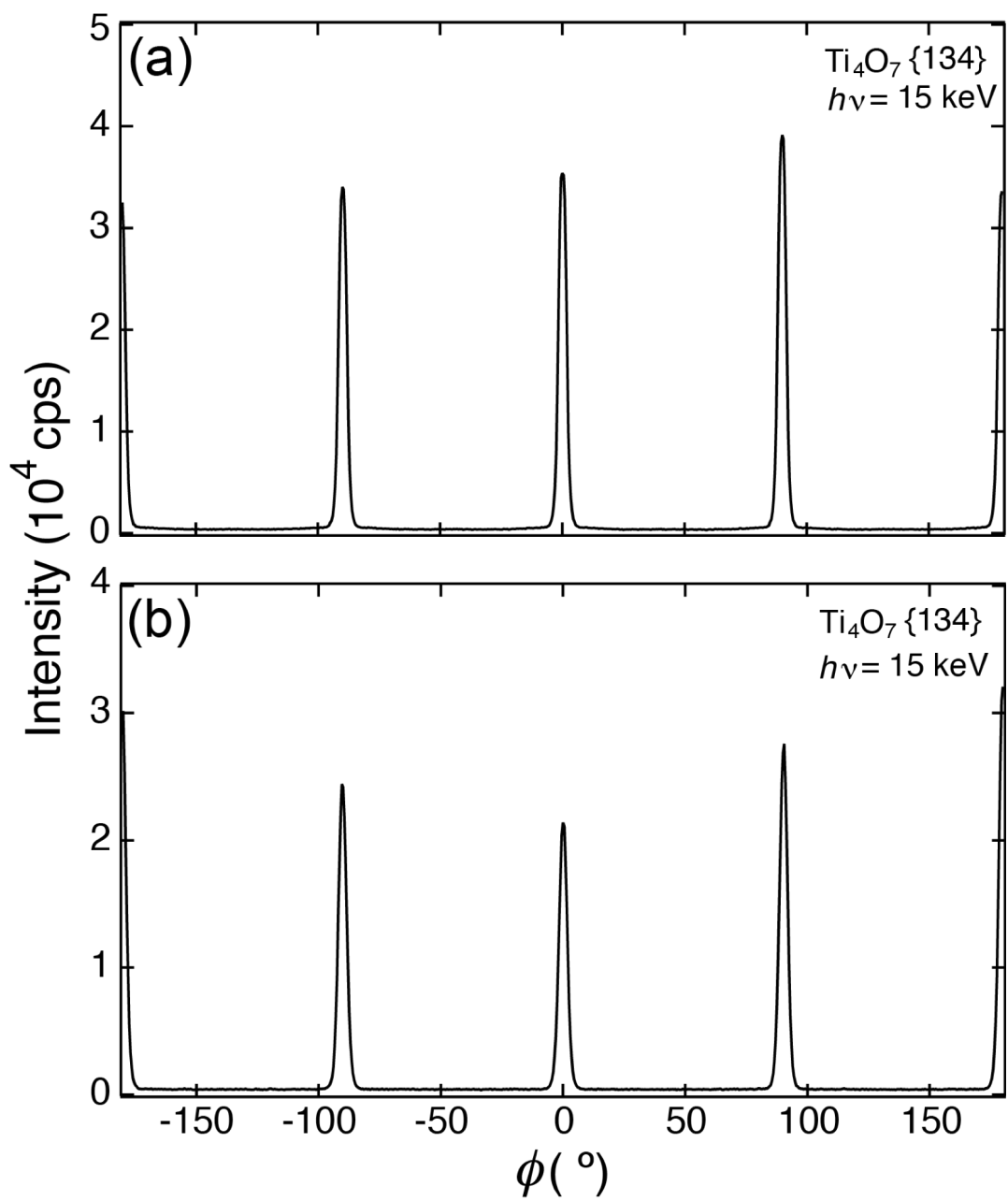

**Figure S4**. XRD azimuth $\phi$-scans of the Ti$_4$O$_7$ 134 reflections for the (a) insulating and (b) superconducting Ti$_4$O$_7$ films.

**XRD measurements for the γ-Ti$_3$O$_5$ film**

Figure S5 shows contour map of film- and substrate-reflections profiles intensity, plotted against scattering angle 2$\theta$ and tilt angle $\chi$, for the γ-Ti$_3$O$_5$ film grown on α-Al$_2$O$_3$ (0001) substrates. The *d* values of interplanar spacing distances, $\chi$ angles obtained by synchrotron radiation XRD measurements (Figs. S6), and corresponding the Miller indices are listed in Table S2. For the γ-Ti$_3$O$_5$ 143 reflection, the XRD azimuth $\phi$-scan around the film normal was also carried out (Fig. S7). The reflections appeared every 60°, indicating the six-fold rotational domains in the film. The in-plane (out-of-plane) orientation relationships were determined to be γ-Ti$_3$O$_5$ [100] // α-Al$_2$O$_3$ [10–10], [01–10], [–1100] and γ-Ti$_3$O$_5$ [–100] // α-Al$_2$O$_3$ [10–10], [01–10], [–1100] (γ-Ti$_3$O$_5$ [011] // α-Al$_2$O$_3$ [0001]).

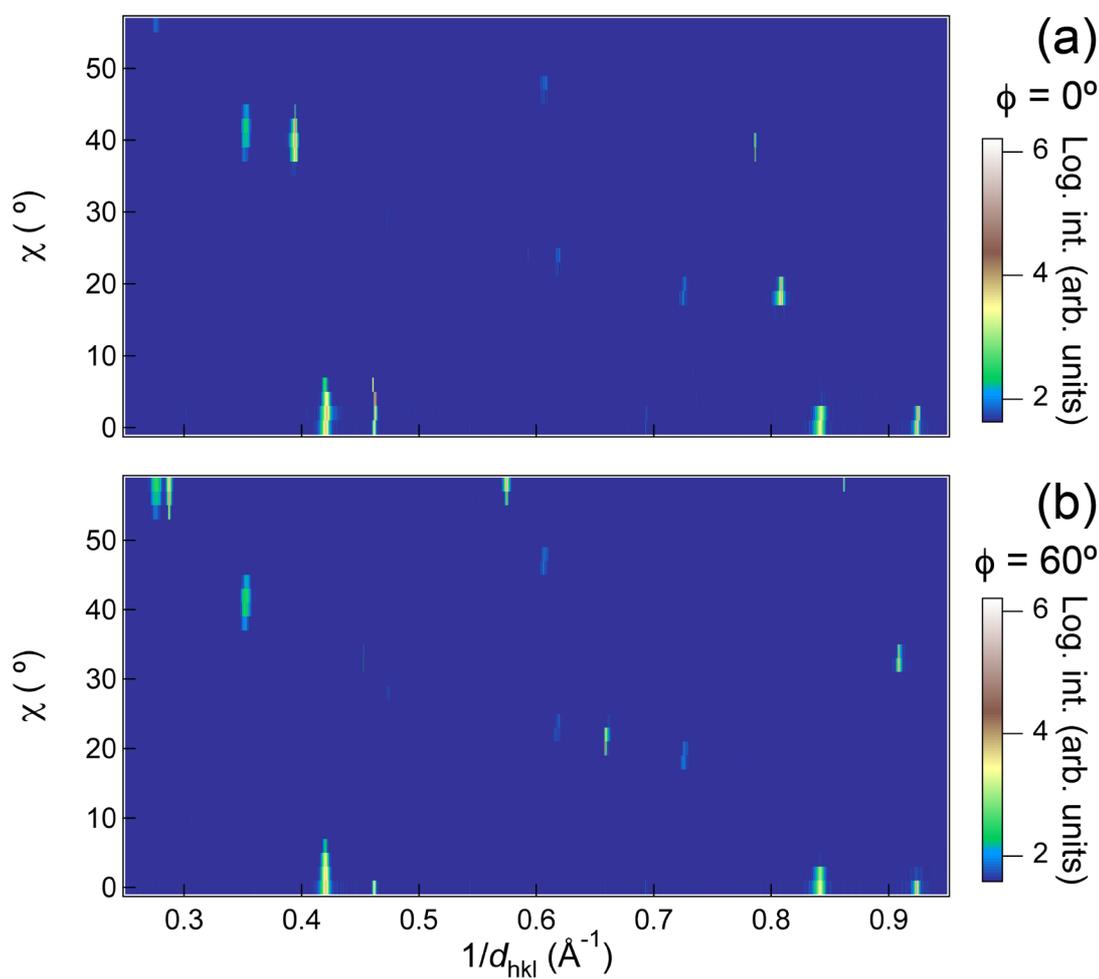

**Figure S5**. Contour map of film- and substrate-reflections intensity, constructed from 2$\theta$-$\theta$ XRD measured by stepwisely varying tilt angle $\chi$ while fixing azimuth angle (a) $\phi = 0°$ and (b) $\phi = 60°$ for the $\gamma$-Ti$_3$O$_5$ film grown on $\alpha$-Al$_2$O$_3$ (0001) substrates.

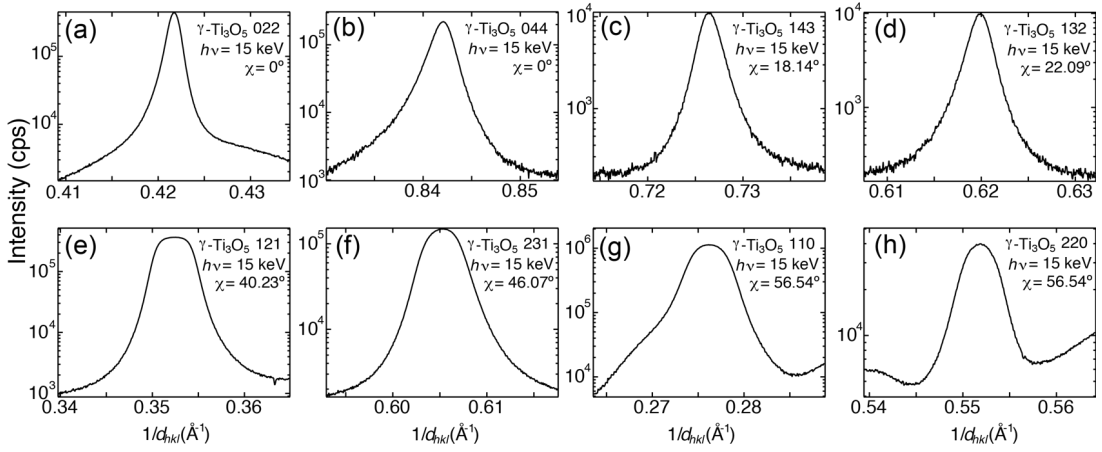

**Figure S6**. Some of the film-reflection profiles measured for the $\gamma$-Ti$_3$O$_5$ film. (a) 022, (b) 044, (c) 143, (d) 132, (e) 121, (f) 231, (g) 110, and (h) 220 reflections.

**Table S2**. List of Miller indices, $d$ values of interplanar spacing distances, and tilt angle $\chi$ for the $\gamma$-Ti$_3$O$_5$ film.

| No | hkl | $d_{hkl}$ (Å) | $\chi$ (°) |
| --- | --- | --- | --- |
| 1 | 022 | 2.375 | 0 |
| 2 | 044 | 1.188 | 0 |
| 3 | 143 | 1.377 | 18.14 |
| 4 | 132 | 1.613 | 22.09 |
| 5 | 121 | 2.839 | 40.23 |
| 6 | 231 | 1.652 | 46.07 |
| 7 | 110 | 3.620 | 56.54 |
| 8 | 220 | 1.812 | 56.54 |

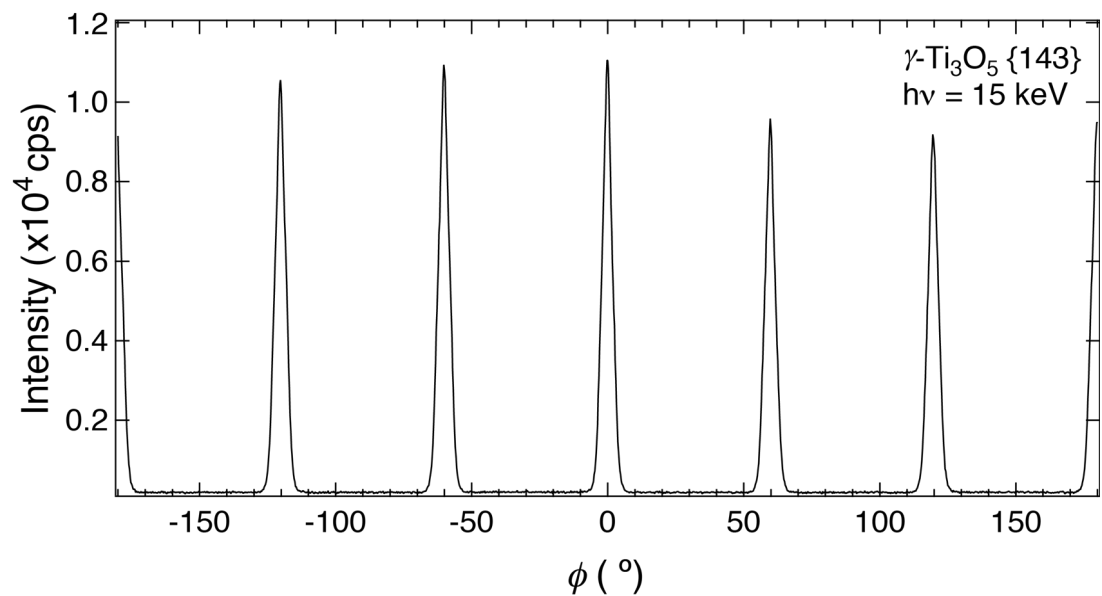

**Figure S7**. XRD azimuth $\phi$-scans of the $\gamma$-Ti$_3$O$_5$ 143 reflections.

**Superconducting volume fraction of $\gamma$-Ti$_3$O$_5$ films**

The diamagnetic signals along with the Meissner effect are usually overestimated for the thin film samples owing to the shape magnetic anisotropy, the signals are analysed by applying a model described as follows. An equation utilized for a uniformly magnetized ellipsoidal sample is given by $M = \frac{\chi H_0}{1+4\pi\upsilon\chi}$, where $M$, $\chi$, $H_0$, and $\upsilon$ are magnetization, magnetic susceptibility, magnetic field, and diamagnetic factor, respectively. The diamagnetic factor $\upsilon$ approximately equals to $1 - \pi d/2R$ ($d \ll R$), where $d$ and $R$ are thickness of the sample and radius of the ellipsoid, respectively. By substituting the experimental values to all the parameters, superconducting volume fraction of the $\gamma$-Ti$_3$O$_5$ film was estimated as shown in Fig. S8. More than 90 ％ of the volume fraction was obtained at 3.3 K, suggesting bulk superconductivity in the $\gamma$-Ti$_3$O$_5$ film.

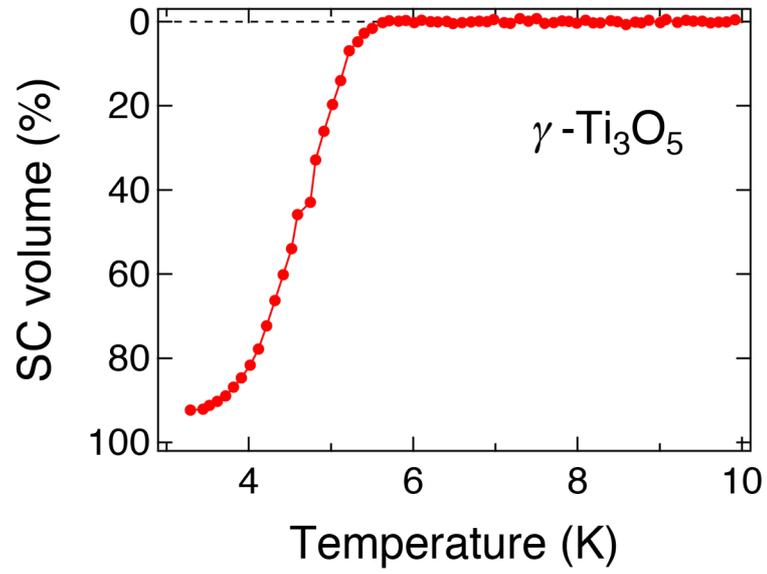

**Figure S8.** Superconducting (SC) volume fraction of the γ-Ti$_3$O$_5$ film estimated from the temperature dependence of magnetization shown in the inset of Fig 4b.

**Superconductivity in $Ti_4O_7$ films grown under $P_{Ar} = 1 \times 10^{-6}$ Torr**

Figure S9 shows temperature dependence of resistivity for the $Ti_4O_7$ film grown under $P_{Ar} = 1 \times 10^{-6}$ Torr. The $P_{O2}$ (residual oxygen gases) in the chamber is expected to be in an intermediate range between those for the growth of insulating ($P_{O2} = 1 \times 10^{-7}$ Torr) and superconducting ($P_{Ar} = 1 \times 10^{-3}$ Torr) films. Clear hysteresis was found at around 150 K [Fig. S9 (a)], corresponding to the metal–insulator transition (MIT) in the normal state. In addition, the superconducting state was also found at low temperatures [Fig. S9 (b)]. $T_{C,\,onset}$ of 2.9 K was slightly lower than that described in the main text. The emergence of the MIT and superconductivity in a sample supports bipolaronic mechanism in the $Ti_4O_7$ film.

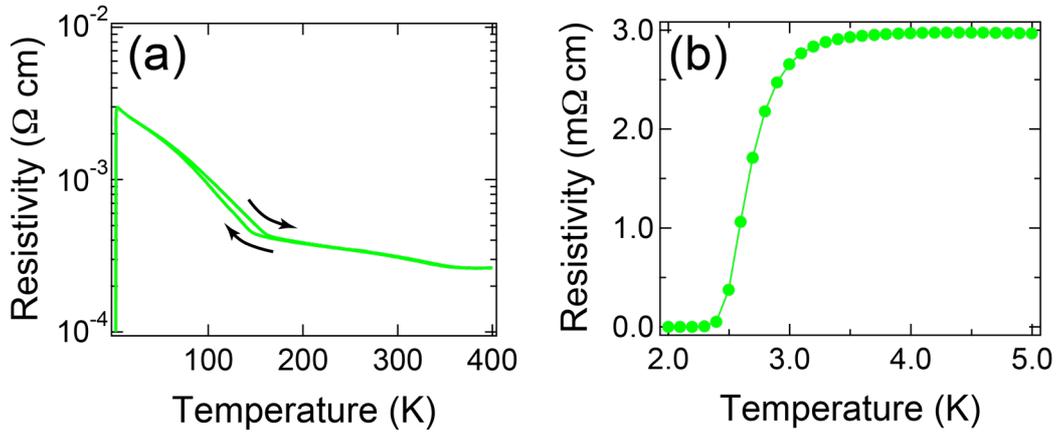

**Figure S9.** Temperature dependence of resistivity for the $Ti_4O_7$ film grown under $P_{Ar} = 1 \times 10^{-6}$ Torr (a) in the whole and (b) a low-temperatures range.

**Superconductivity in Ti$_4$O$_7$ films grown on MgAl$_2$O$_4$ (100) substrates**

Figure S10 shows temperature dependence of resistivity for the Ti$_4$O$_7$ film grown on MgAl$_2$O$_4$ (100) substrate. The resistivity curve was in good agreement with that grown under $P_{Ar} = 1 \times 10^{-3}$ Torr on LSAT (100) substrate. Superconductivity was clearly observed at low temperatures.

Crystal structure of the film was investigated by XRD with Cu K$\alpha_1$ radiation. Figure S11(a) shows 2$\theta$-$\theta$ XRD patterns of the film with various tilt angle $\chi$. For comparison, 2$\theta$-$\theta$ XRD patterns of the Ti$_4$O$_7$ film grown on LSAT (100) substrate were shown in Fig. S11(b). The film reflections were found at the similar angles, indicating that the film on MgAl$_2$O$_4$ (100) substrate was out-of-plane (101)-oriented Ti$_4$O$_7$. Emergence of superconductivity of Ti$_4$O$_7$ films on the different substrates confirms that superconducting phase at low temperatures is intrinsic to the Ti$_4$O$_7$ films themselves. Furthermore, superconductors composed of Mg, Al, Ti, and O with $T_c$ of more than 3 K are not yet known.

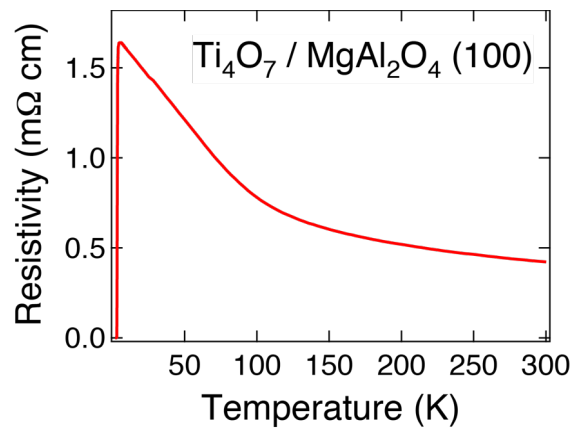

**Figure S10.** Temperature dependence of resistivity for Ti$_4$O$_7$ film grown on MgAl$_2$O$_4$ (100) substrate.

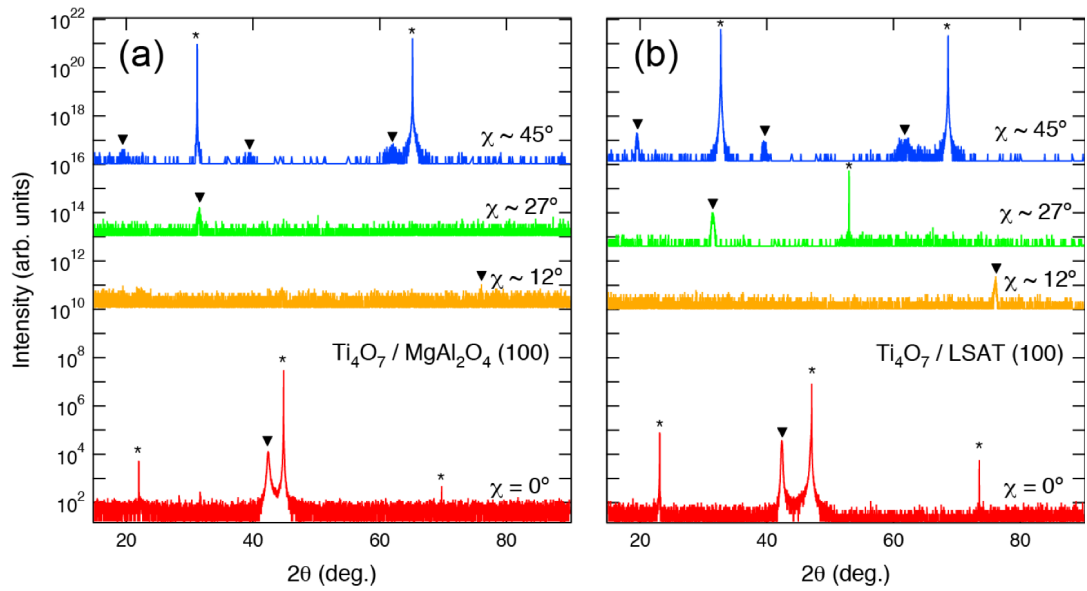

**Figure S11.** $2\theta$-$\omega$ XRD patterns of Ti$_4$O$_7$ films grown on (a) MgAl$_2$O$_4$ (100) and (b) LSAT (100) substrates with various tilt angle $\chi$. The asterisks and triangles indicate the substrates and films reflections, respectively.